\documentclass{jltp}

\newcommand{\beq}{\begin{equation}}
\newcommand{\eeq}{\end{equation}}
\newcommand{\bea}{\begin{eqnarray}}
\newcommand{\eea}{\end{eqnarray}}
\newcommand{\etal}{{\em et al.}}

\newcommand{\RR}{{\bf R}}
\newcommand{\rr}{{\bf r}}
\newcommand{\PP}{{\mathcal{P}}}
\newcommand{\fwidth}{4in}

\usepackage{graphicx} 

\title{{\scriptsize \sf \vspace*{-50pt}Proceedings article of the 5th Conference on Cryocrystals and Quantum Crystals 
in Wroclaw, Poland,\\[-8pt] submitted to {\em J. Low. Temp. Phys.} (2004).}\\[30pt]
Hydrogen-Helium Mixtures at High Pressure}

\author{Burkhard Militzer}

\address{Geophysical Laboratory, Carnegie Institution of Washington, \\
5251 Broad Branch Road, NW, Washington, DC 20015, USA}

\runninghead{B. Militzer}{Hydrogen-Helium Mixtures at High Pressure}

\begin{document}

\maketitle

\begin{abstract}
The properties of hydrogen-helium mixtures at high pressure are
crucial to address important questions about the interior of Giant
planets e.g. whether Jupiter has a rocky core and did it emerge via
core accretion? Using path integral Monte Carlo simulations, we study
the properties of these mixtures as a function of temperature, density
and composition. The equation of state is calculated and compared to
chemical models. We probe the accuracy of the ideal mixing
approximation commonly used in such models. Finally, we discuss the
structure of the liquid in terms of pair correlation functions.

\end{abstract}

\section{INTRODUCTION}

Hydrogen and helium are the two most abundant elements in giant
planets. While Jupiter and Saturn are well characterized on the
surface, many basic questions about its interior have not been
answered\cite{Gu02}. Jupiter's surface composition has been measured
{\em in situ} by the Galileo probe:\cite{FH03} H 74.2\% by weight, He
23.1\%, and 0.027\% heavier elements, which is enhanced compared to
the protosolar composition: 0.015\% heavier elements along with H
73.6\% and He 24.9\%. The abundance of heavy element in the interior
and their distribution are not well characterized.\cite{SG04} In
particular, it is conversial whether Jupiter has a rocky core. The
detection of a core in Jupiter may validate the standard model of
giant-planet formation, nucleated capture of nebular
hydrogen.\cite{Li93} An alternative scenario was proposed by Alan Boss
who suggested that giant planets form directly from spiral
instabilities in protostellar disks.\cite{Boss97,Boss04} Under this
gravitational instability hypothesis, giant planets would not have a
core, or at least a much smaller one.

Since there is no direct way to detect a core in Jupiter, one must
instead refer to indirect measurements and to models for the planet
interior. Such models are constrained by the available observation
data, in particular, the properties at the planet surface and the
gravitational moments measured through fly-by trajectories. All models
rely on an equation of state (EOS) of hydrogen-helium
mixtures. However, the uncertainties in the available EOS are large
and have not allowed one, among many other questions, to determine
whether Jupiter has a core.

In this article, we present results from path integral Monte Carlo
(PIMC) simulations\cite{Ce95} that enable us to study quantum many-body
systems at finite temperature from first principles. In this
simulation, hydrogen-helium mixtures are represented by an ensemble of
electrons, protons and helium nuclei, each described by a path in
imaginary time to incorporate quantum effects. The electrons are
treated as fermions while exchange effects for the nuclei can be
neglected for the considered thermodynamics conditions.

\section{PATH INTEGRAL MONTE CARLO}

The thermodynamic properties of a many-body quantum system at finite
temperature can be computed by averaging over the density matrix,
$\hat{\rho} = e^{-\beta \hat{H}}, \beta=1/k_{\rm B} T$.  
Path integral formalism\cite{Fe53} is based on the identity,
\beq
e^{-\beta \hat{H}} = \left[ e^{-\frac{\beta}{M} \hat{H}} \right]^M
\eeq
where $M$ is a positive integer. Insertion of complete sets of states
between the $M$ factors leads to the usual imaginary time path
integral formulation, written here in real space,
\beq
\rho(\RR,\RR';\beta)=
\int\ldots\int d\RR_{1}\ldots d\RR_{M-1} \; \rho(\RR,\RR_{1};\tau)\ldots\rho(\RR_{M-1},\RR';\tau)
     \label{eq2.5}
\eeq
where $\tau=\beta/M$ is the time step. Each of the $M$ steps in the
path now has a high temperature density matrix
$\rho(\RR_k,\RR_{k+1};\tau)$ associated with it. The integrals are
evaluated by Monte Carlo methods. 
The density matrix for bosonic and fermionic systems can be obtained
by projecting out states of corresponding symmetry. In PIMC, one sums
up different permutations $P$,
\beq
\nonumber
\rho_{\rm b/f}(\RR,\RR';\beta) 
= 
\frac{1}{N!} \sum_\PP \; (\pm 1)^\PP \rho(\RR,\PP \RR' ; \beta)
= \frac{1}{N!} \sum_\PP \; (\pm 1)^\PP \; \! \! \! \! \! \! \! \int \limits_{\RR \rightarrow \PP \RR' } 
\! \! \! \! \! \! d\RR_t \, e^{-U[\RR_t] }.
\eeq
For bosons, this is essentially an exact numerical algorithm that has
found many applications.\cite{Ce95} For fermions, the cancellation of
positive and negative contributions leads to numerically unstable
methods, which is known as the {\em fermion sign problem}. Ceperley
showed that this problem can be solved by introducing the {\em
restricted path} approximation,\cite{Ce91,Ce96}
\beq 
\rho_f(\RR_0, \RR' ;\beta) \approx
\frac{1}{N!}\; \sum_\PP \; (-1)^\PP 
\int\limits_{\scriptsize {
\! \! \! \! \! \! \! \! \! \! \! \! \! \!\! \! \! \! 
\begin{array}{c} 
  \RR_0 \rightarrow \PP \RR'\\
  {\rho_T(\RR(t),\RR_0;t)>0}
\end{array}}
 \! \! \! \! \! \! \! \! \! \! \! \! \! \! \! \!
\! \! \! \! \! \! \! \! \! \! } d\RR_t \;\; e^{-U[\RR(t)] } \quad,
\label{restricted_PI} 
\eeq
where one only samples path $\RR(t)$ that stay within the positive
region of a {\em trial} density matrix,
$\rho_T(\RR(t),\RR_0;t)>0$. This procedure leads to an efficient
algorithm for fermionic systems. All negative contributions to
diagonal matrix elements are eliminated.\cite{MPC03} Contrary to the
bosonic case, the algorithm is no longer exact since it now depends on
the approximations for the trial density matrix. However, the method
has worked very well in many applications.\cite{PC94,MP02} For
$\rho_T$, one can use the free particle or a variational density
matrix.\cite{MP00}
In the results presented here, the
high temperature density matrix was taken as a product of exact pair
density matrices,
\bea
\frac{\rho(\RR,\RR';\tau)}{\rho_0(\RR,\RR';\tau)} 
&=& 
\left< e^{-\int_0^\tau dt \sum_{i<j} V(\rr_{ij}) } \right>_{\rm \RR \to \RR'}
=
\left< \prod_{i<j} e^{-\int_0^\tau dt V(\rr_{ij}) } \right>_{\rm \RR \to \RR'}\\
&\approx&
\prod_{i<j} \left< e^{-\int_0^\tau dt V(\rr_{ij}) } \right>_{\rm \rr_{ij} \to \rr_{ij}'} 
\equiv
e^{ - \sum_{i<j} u(\rr_{ij},\rr_{ij}'; \tau) },
\eea
where $\rho_0(\RR,\RR';\tau)$ is the free particle density matrix and
$u(\rr_{ij},\rr_{ij}'; \tau)$ is the pair action for paths initially
separated by $\rr_{ij}$ and finally at time $\tau$ by $\rr'_{ij}$. An
approximation is introduced by assuming that the different pair
interactions can be averaged by independent Brownian random walks that
are denoted by brackets $\left<\ldots\right>$. This approach is
efficient but not exact, and therefore puts a limit on the imaginary
time step $\tau$ in many-body simulations. The pair action, $u$, can
be computed by different methods.\cite{St68,Po88,KS95}
The following simulation results were derived with $N_e=32$ electrons
and the corresponding number of protons, $N_p$, and helium nuclei
$N_{\rm He}$ to obtain a neutral system ($N_e=N_p+2 N_{\rm He}$) with
helium fraction $x\equiv 2 N_{\rm He}/N_e$. Periodic boundary conditions
are applied. As imaginary time discretization, we employ
$\tau=0.079$. (Atomic units of Bohr radii and Hartrees will be used
throughout this article.)

The electrons are treated as fermions with fixed spin. We use
variational nodes to restrict the paths and have therefore extended
the approach in Ref.~[\onlinecite{MP00}] to mixtures. The nuclei
are also treated as paths but their exchange effects are not relevant
here.

\section{PHASE DIAGRAM OF HOT DENSE HYDROGEN}

\begin{figure}[htb]
\centerline{\includegraphics[angle=0,width=\fwidth]{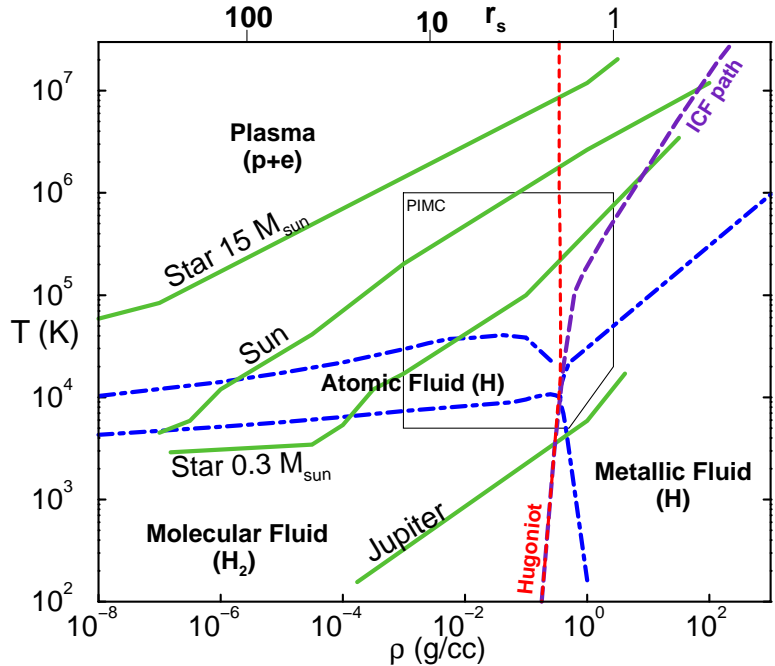}}
\caption{Density-temperature phase diagram of hot dense hydrogen. 
       The blue dash-dotted lines separate the molecular, the atomic,
       the metallic, and the plasma regime. The green solid lines are
       isentropes for Jupiter and stars with 0.3, 1, and 15 solar
       masses. Single shock Hugoniot states as well as the inertial
       confinement fusion path\cite{lindl} are indicated by dashed
       lines. The thin solid line show $\rho$-$T$ conditions of PIMC
       simulations.}
\label{phase}
\end{figure}

Figure~\ref{phase} shows the high temperature phase diagram of dense
hydrogen beginning with the fluid and reaching up to a highly ionized
plasma state. The figure includes the isentropes for Jupiter and low
mass stars\cite{SC95} and indicates the thermodynamic conditions, at
which PIMC simulations have been applied.\cite{MC00,MC01,MP02} We are
now going to use these simulation results to characterize hot dense
hydrogen from a path integral perspective.

\begin{figure}[tb]
\centerline{\includegraphics[angle=0,width=\fwidth]{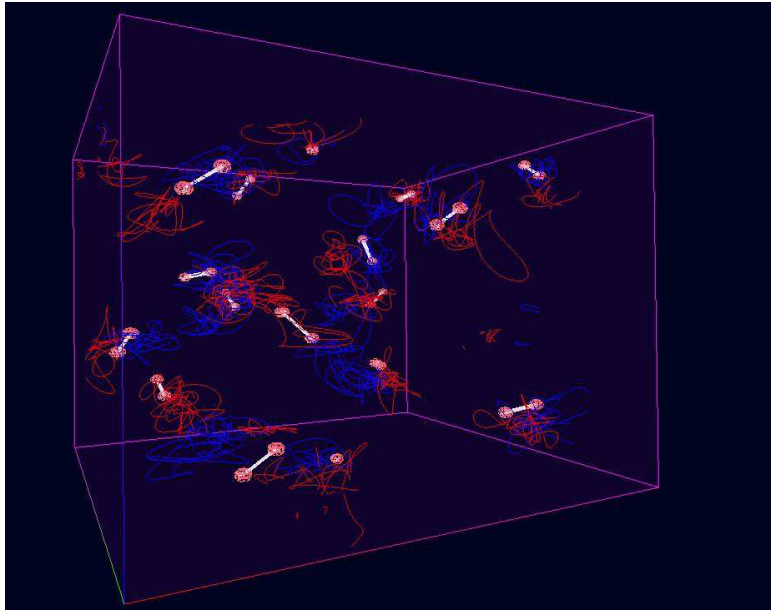}}
\vspace*{3mm}
\caption{Snapshot from PIMC simulations of pure hydrogen in the molecular regime at 
         $T=5\,000\,\rm K$ and $r_s=4.0$. The pink spheres denote the
         protons. The bonds (white lines) were added as a guide to the
         eye. The electron paths are shown in red and blue [light and
         dark gray] depending on their spin state. }
\label{pict_dilute_H2}
\end{figure}

At low density and temperature, one finds a fluid of interacting
hydrogen molecules. A PIMC snapshot for $T=5\,000\,\rm K$ and
$r_s=4.0$ is shown in Fig.~\ref{pict_dilute_H2}. The proton paths are
very localized due to their high mass. Their spread can be estimated
from the de Broglie thermal wavelength, $\lambda_{d}^{2}=\hbar^{2}
\beta /2\pi m$. The electron paths are more spread out but they 
are localized to some extent since two electrons of opposite spin
establish the chemical bond in the hydrogen molecule.

If the temperature is raised from 5000 to $10^6$ K, hydrogen
undergoes a smooth transition from a molecular fluid though an atomic
regime and finally to a two-component plasma of interacting electrons
and protons. Many-body simulations at even higher temperatures are not
needed since analytical methods like the Debye-H{\"u}ckel theory work
well. PIMC with explicit treatment of the electrons can also be
applied to temperatures below 5$\,$000$\,$K but groundstate methods
are then more practical since excitation become less relevant and the
computational cost scale with the length of paths like $T^{-1}$.

A detailed analysis of the chemical species present in the low density
regime ($r_s \ge 2.6$) is given in Ref.~\onlinecite{MC00}. With
decreasing density, one finds that the degree of molecular
dissociation increases since the atomic state has higher entropy. For
the same reason, one observes that the degree of atomic ionization
increases with decreasing density. All these low-density effects can
be well characterized by analytical models based on approximate free
energy expressions for atoms, molecules and ionized
particles.\cite{SC92,Ro90,rotesbuch,Ju01} In this regime, one also
finds reasonably good agreement between for the EOS derived from PIMC
and chemical models.\cite{MC00}

\begin{figure}[tb]
\centerline{\includegraphics[angle=0,width=\fwidth]{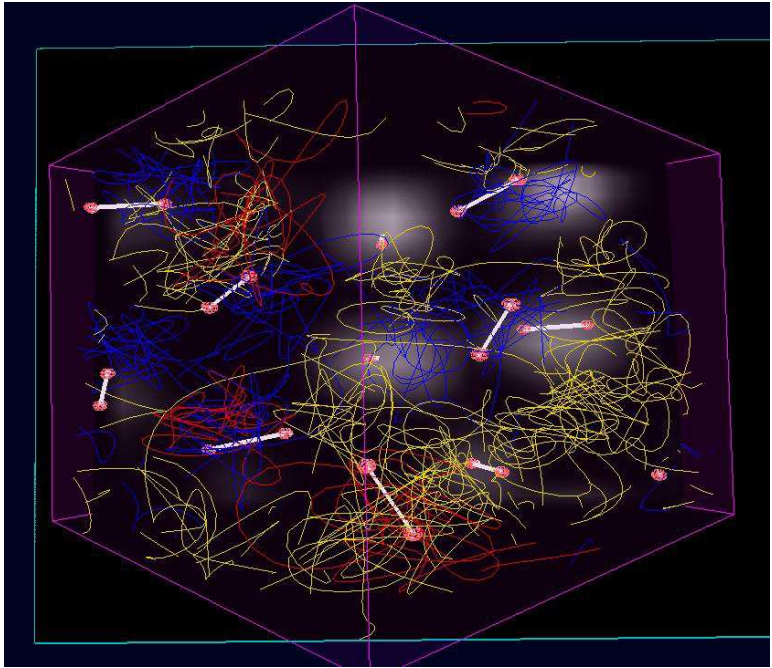}}
\caption{Deuterium in the dense molecular regime ($T=5\,000\,\rm K$ and $r_S=1.86$) 
       Due to the density increase compared to
       Fig.~\ref{pict_dilute_H2} (see details there), the electron
       paths permute with a rising probability (shown as yellow [light
       gray] lines) but are still localized enough to form a bond
       between the two protons in the molecule. The electron density
       average over many electron configurations is indicated in gray
       color on the blue rectangles.}
\label{pict_dense_H2}
\end{figure}

If the density is increased at $T=5\,000\,$K, one finds an
intermediate regime of strongly interacting molecules
(Fig.~\ref{pict_dense_H2}). Some electron paths exchange with
neighboring molecules indicating the importance of fermionic
effects. However, the electrons are still localized enough to provide
a sufficient binding force for the protons.

\begin{figure}[tb]
\centerline{\includegraphics[angle=0,width=\fwidth]{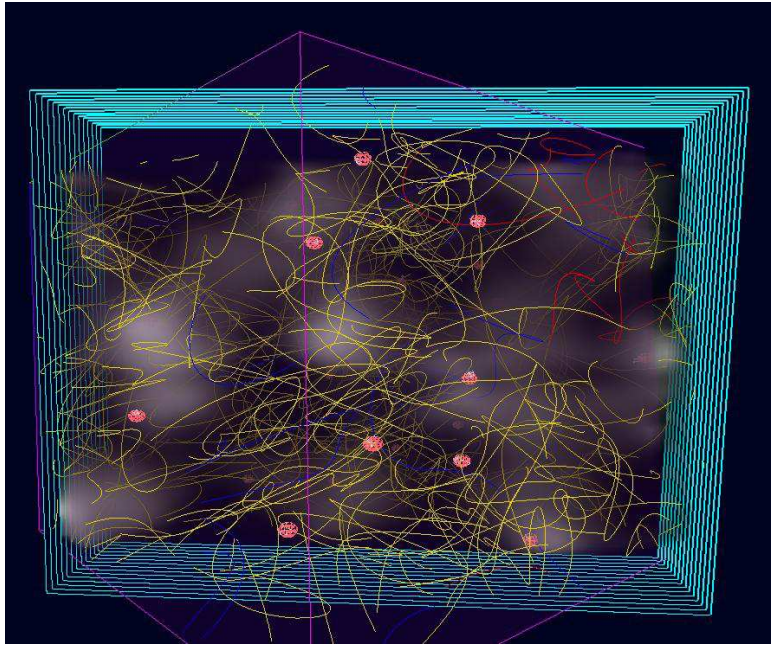}}
\caption{Deuterium in the metallic regime characterized by unpaired 
	protons and a gas of degenerate electrons. The snapshot was
	taken from a PIMC simulation at $T=6\,250\,\rm K$ and
	$r_s=1.60$. The electron paths are delocalized and exchange
	frequently (see description of Fig.~\ref{pict_dense_H2}).}
\label{pict_metallic_H}
\end{figure}

If the density is increased further from $r_s=1.86$ to 1.60, this
binding force is lost due to further delocalization of the electrons
(Fig.~\ref{pict_metallic_H}). Almost all electrons are now involved in
long exchange cycles indicating a highly conducting, metallic
state. No binding forces of the protons can be observed. One can
therefore conclude that hydrogen metallizes in dissociated, or atomic
form, a conclusion that is consistent with DFT
simulations\cite{Banier2000,Scandolo2003,Desjarlais2003,Bonev2004}. If
the density is increased further, the electrons form a rigid
neutralizing background and one recovers the limit of a one-component
plasma of protons. If the temperature is increased, the electron paths
get shorter and shorter, the degree of electron degeneracy decreases
gradually and one recovers the limit of a two-component plasma at high
temperature.

The nature of the molecular-metallic transition has not yet been
determined with certainty. A large number of analytical
models\cite{Be99} predict a first order transition, others do
not.\cite{Ro90} PIMC simulation with free particle nodes by
W. Magro~{\em et al.}\cite{Ma96} showed an abrupt transition characterized by
a region $\left.\frac{dP}{dT}\right|_V<0$ for $r_s$=1.86. Later work
by B. Militzer\cite{BM00} using more accurate variational nodes did
not show such a region for $r_s$=1.86. Whether PIMC simulations with
variational nodes predict a gradual molecular-metallic transition for
all temperatures, or if the region of an abrupt transition has shifted
to lower, not yet accessible, temperatures has not been
determined. However, recent density functional molecular dynamics
simulations by various authors predict a first order transition below
$5\,000\,$K.\cite{Banier2000,Scandolo2003,Desjarlais2003,Bonev2004}

There was a lot of recent interest focused on the hydrogen EOS because
of the {\em unexpectedly high compressibility} inferred from the
laser-driven shock wave experiment by Da Silva~\etal\cite{Si97} and
Collins~\etal\cite{Co98} using the Nova laser at Lawrence Livermore
National Laboratory. The measurements indicated that hydrogen could be
compressed to about 6-fold the initial density rather than 4-fold as
indicated by the Sesame model.\cite{Ke83,Kerley2003} Even though the
temperatures and pressures reached in shock experiments are higher
than those in the giant planets these measurements are crucial here
since they represent the only way to distinguish between different EOS
models at high temperature.

The Nova results challenged the existing understanding of high P-T
hydrogen and triggered many new experimental and theoretical
efforts. Different chemical models gave rise to very different
predictions\cite{Be99,Ro90,Ju00} ranging from 4-fold compression as
suggested in\cite{Ke83,Kerley2003} to 6-fold as predicted by the Ross
model.\cite{Ro98} While the accuracy of chemical models did not allow
any conclusive predictions, first principle simulations from
PIMC\cite{MC00,Mi01} as well as from density functional molecular
dynamics\cite{Le00,Desjarlais2003,Bonev2004} consistently predict a
lower compressibility of about 4.3.

Since then there have been many attempts to resolve this discrepancy
but the most important contributions came from new experiments by
Knudson~{\em et al.}\cite{Kn01,Kn03,Kn04} at Sandia National
Laboratory. Instead of a laser drive, they used magnetically driven
shock waves in combination with bigger samples. They found a
significantly lower compressibility quite similar to predictions from
first principles methods. The new results are also supported by a
third set of experiments by Russian investigators using spherically
converging shock waves.\cite{Belov2002,Boriskov2003}

\section{HYDROGEN-HELIUM MIXTURES}

Due to importance for astrophysical applications, EOS models for
hy\-dro\-gen-he\-lium mixtures have been studied for quite some time. The
most widely used EOS was derived by Saumon, Chabrier and van
Horn (SCH).\cite{SC95} Like previous chemical models, it is based on a
hydrogen and a helium EOS that combined using an ideal mixing
rule. 

Following the discussion of the hydrogen EOS above, we are now
analyzing the second ingredient: the EOS of helium. Given its low
abundance in giant planets, helium is not expected to be present in
very high concentration. However,
Stevenson\cite{Stevenson77b} has proposed
that the hydrogen-helium mixture could phase separate under certain
high pressure conditions. As a result, helium droplets would form and
fall as rain towards planet core. The associated release of latent
heat would delay the cooling process of the planet. This was suggested
as one possible explanation of why standard models predict Saturn to cool
at a much faster rate than is observed.\cite{FH04} In order to
detect hydrogen-helium phase separation one needs an EOS of pure
helium that we will now discuss.

\begin{figure}[htb]
\centerline{\includegraphics[width=2.48in]{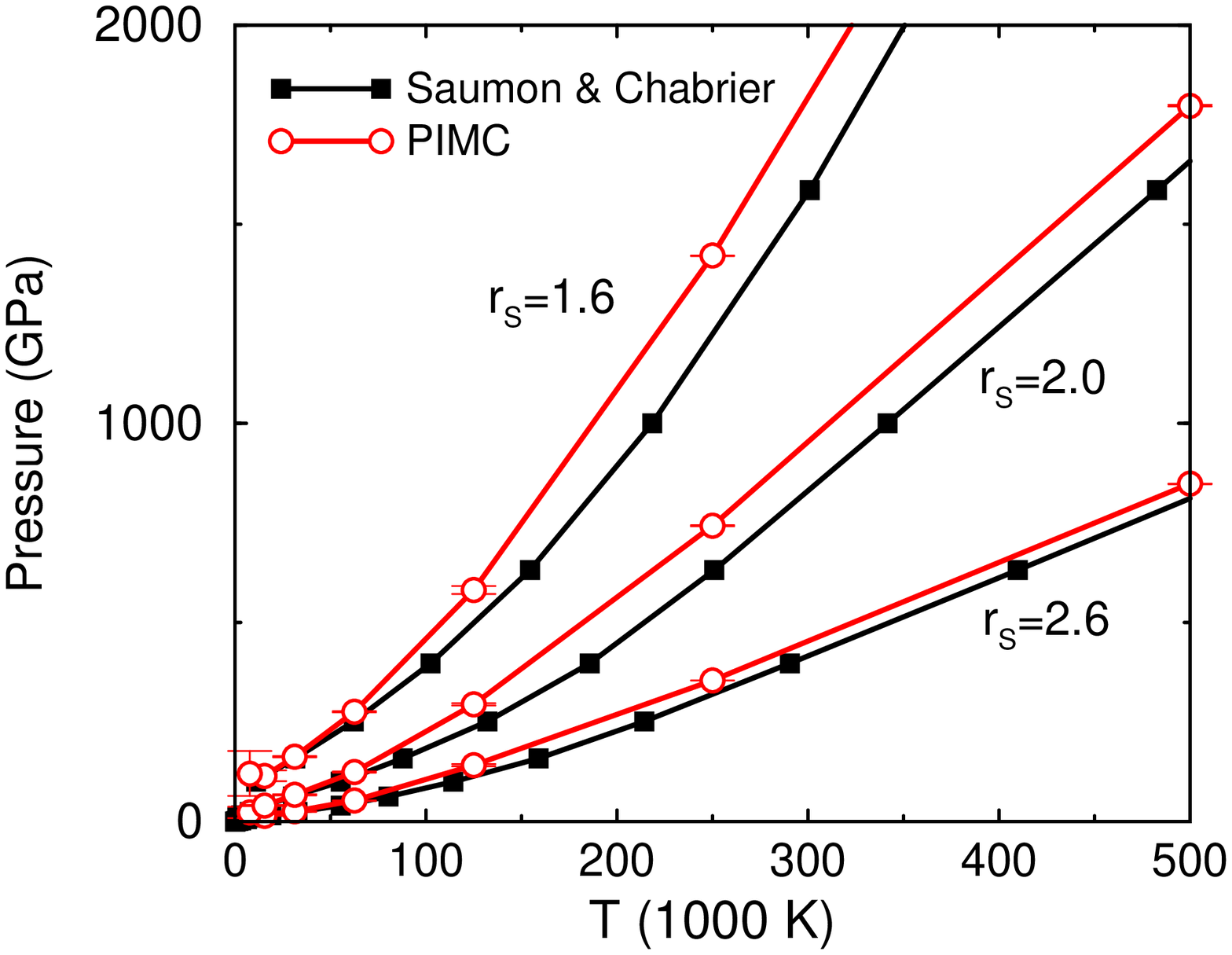}~~~~\includegraphics[width=2.4in]{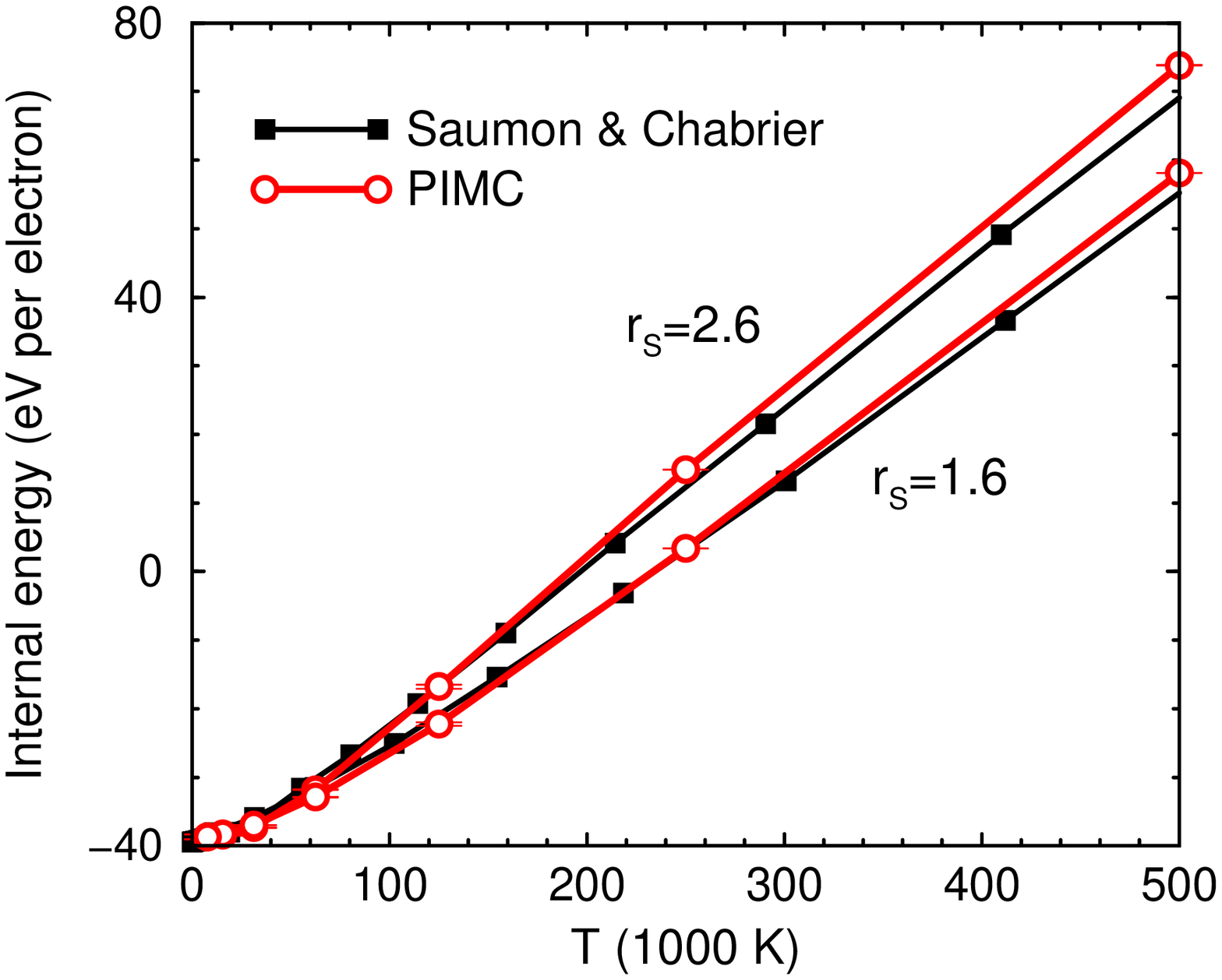}}
\caption{Equation of state comparison for pure helium showing the pressure (left) 
       and the internal energy (right) for different density as
       derived from path integral Monte Carlo simulations and the
       chemical model by Saumon, Chabrier, and van Horn.\cite{SC95}}
\label{HeEOS}
\end{figure}

Figure~\ref{HeEOS} shows an EOS comparison of our PIMC calculations
with the SCH model. One finds that the pressure is underestimated by
SCH and that the deviations increase with density. The internal energy
at high temperature is also underestimated. The analysis suggests a
more careful treatment of the helium ionization states is needed to
improve the chemical model.

To conclude the comparison with chemical models, we test one more
assumption that is generally made: the ideal mixing hypothesis. We
performed a large number of PIMC calculations of fully interacting
hydrogen-helium mixtures at a fixed density of $r_s=1.86$ for various
temperatures and mixing ratios. The resulting correction to ideal
mixing is given by,
\beq
\Delta f_{\rm mix} = f(x) - (1-x)\,f(x=0) - x\,f(x=1)\,.
\eeq 

\begin{figure}[htb]
\centerline{\includegraphics[width=2.48in]{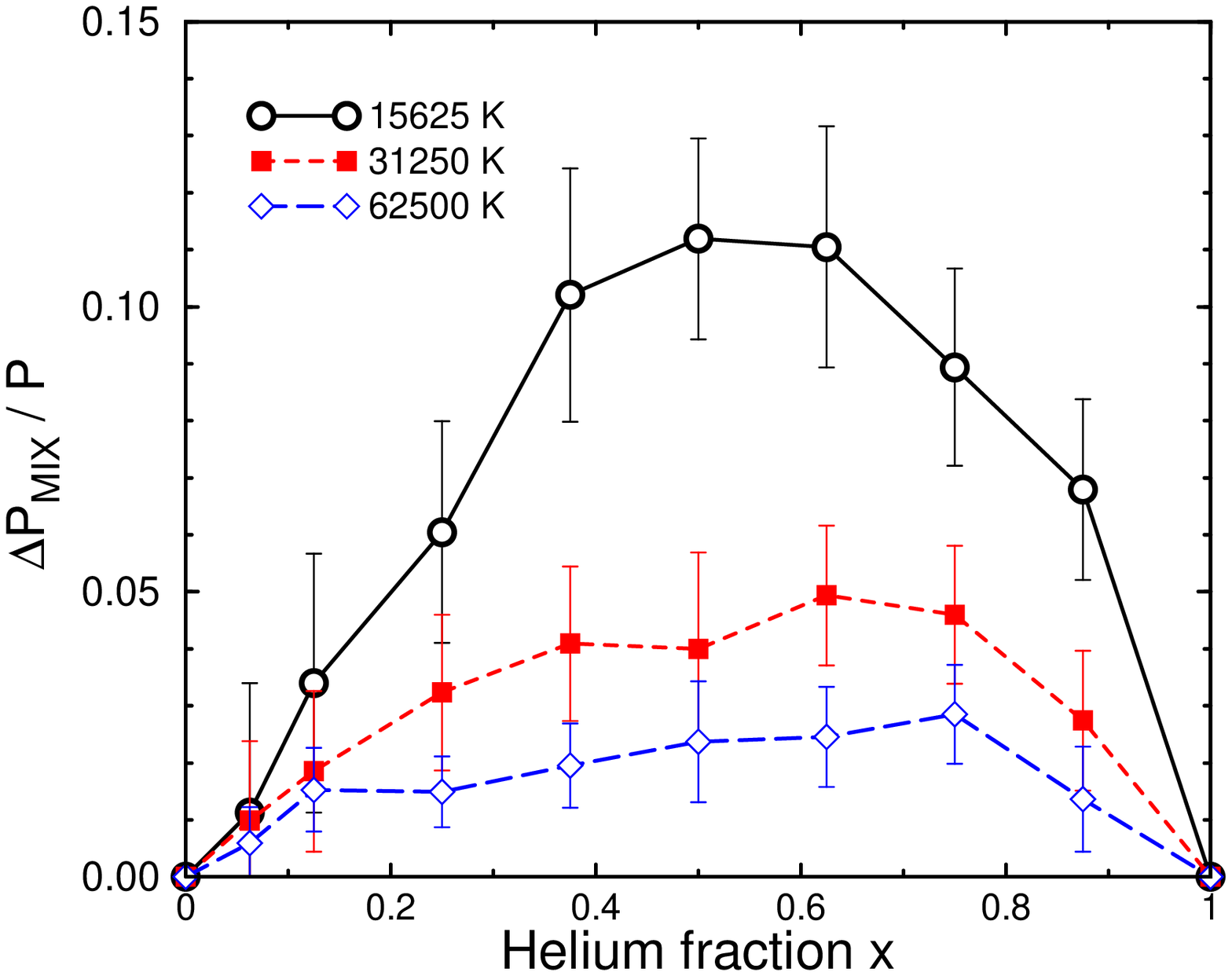}~~~~\includegraphics[width=2.4in]{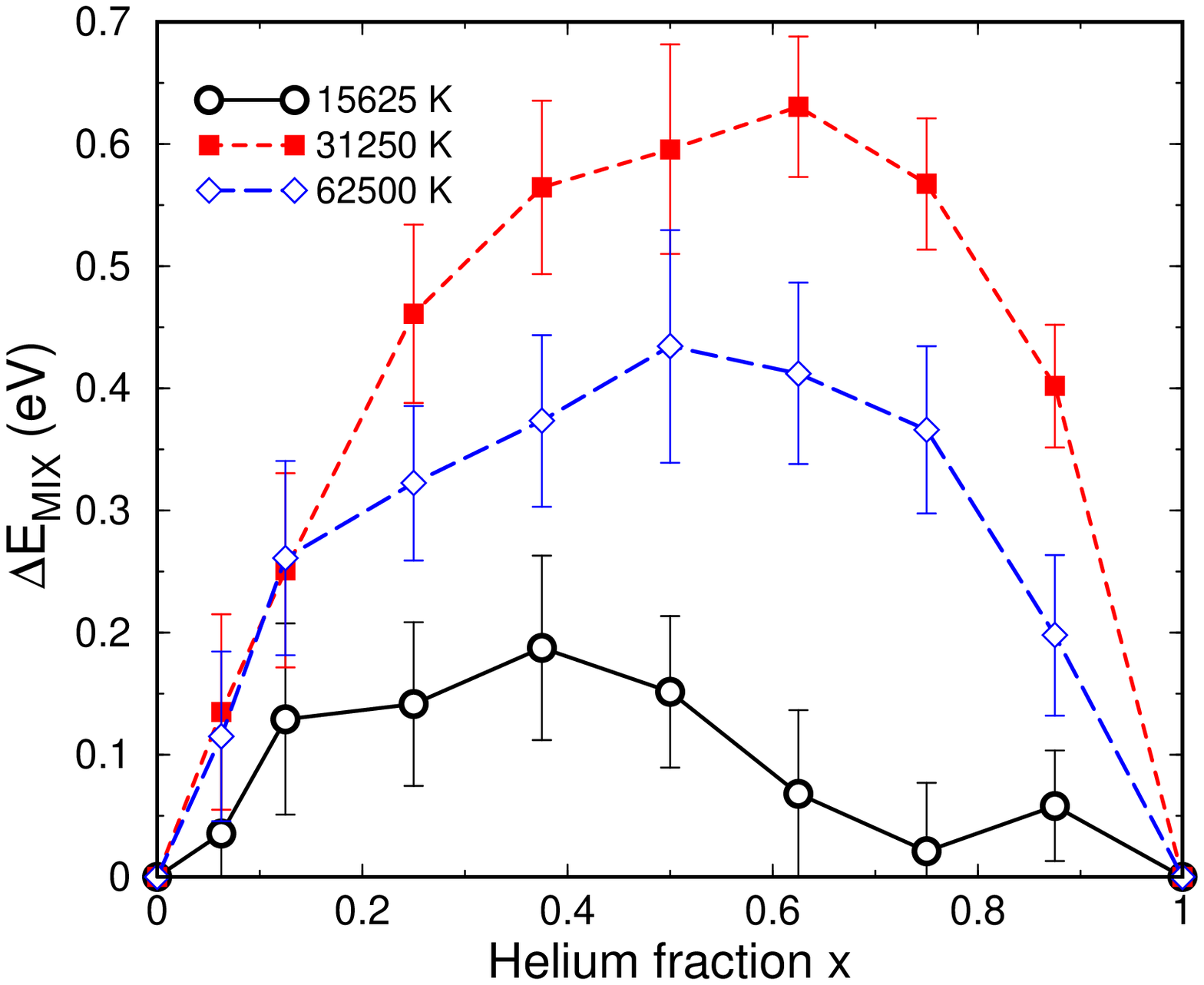}}
\caption{Excess mixing pressure $\Delta P_{\rm mix}$ and internal energy per electron $\Delta E_{\rm mix}$
       are shown for a hydrogen-helium mixture at $r_s=1.86$. Path
       integral Monte Carlo results are shown for several
       temperatures. The mixing is performed at constant
       density. $\Delta E_{\rm mix}$ is largest for T=31$\,$250$\,$K
       because at this temperature the two end members are
       characterized by very different degrees of ionization.}
\label{mix}
\end{figure}

Figure~\ref{mix} shows that the corrections to pressure, $\Delta
P_{\rm mix}$, increase with decreasing temperature reaching 10\% for
$15\,625\,$K. In the considered temperature interval, the corrections
to the internal energy are largest for $31\,250\,$K which can be
explained by the different ionization states of the two fluids that
lead to larger error if ideal mixing is assumed.

\begin{figure}[htb]
\centerline{\includegraphics[width=\fwidth]{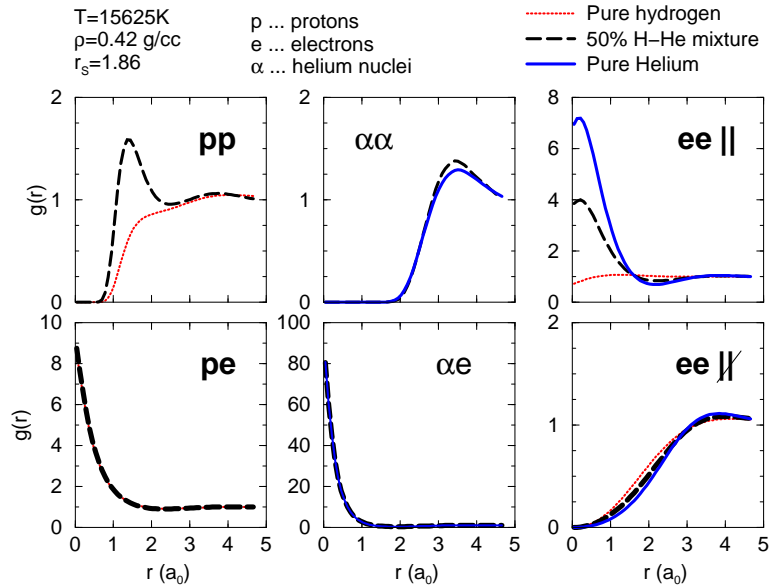}}
\caption{Comparison of different pair correlation functions, $g(r)$, from three PIMC 
       simulations at $r_s=1.86$ and $15\,625\,$K: pure hydrogen (red
       dotted lines), a $x=50\%$ mixture (black dashed lines) and pure
       helium (blue solid lines). The six graphs show $g(r)$ for
       different pairs of protons (p), electrons (e) and helium nuclei
       ($\alpha$). For electrons, pairs with antiparallel (upper
       right) and with parallel (lower right) spins are
       distinguished.}
\label{fig:gr}
\end{figure}

Finally, we discuss pair correlation functions, $g(r)$, for hydrogen-helium mixtures,  
\beq
    g(r)\equiv \frac{\Omega}{N^2}
          \left<\sum_{i\neq j}\delta ({\bf r}-{\bf r}_{ij})
         \right>\;.
                       \label{eq3.1}
\eeq
$g(r)$ functions are a standard tool to characterize the short-range
correlation of particles in liquids. In Fig.~\ref{fig:gr}, we compare
the correlation functions for pure hydrogen, a $x$=50\% mixture, and
pure helium at fixed temperature (15$\,$625$\,$K) and density
($r_s=1.86$). Changes in the pair correlations are now discussed as a
function of helium concentration. The electron-proton $g(r)$ shows a
strong peak at the origin due to the Coulomb attraction. The peak is
not affected when 50\% helium is added to a pure hydrogen
sample. Similarly, the correlation between electrons and helium nuclei
is not altered by the presence of protons.

The electron-electron correlation depends strongly on spin. For pairs
with parallel spin, exchange effects lead to a strong repulsion. The
resulting $g(r)$ function does not depend much on whether helium is
present or not. The correlation function between pairs of antiparallel
spins, on the other hand, is strongly affected by the presence of
helium nuclei. In the helium atom, two electrons with opposite spin
are attracted to the core, which indirectly leads to the observed
increase in the electron-electron $g(r)$.

It is interesting to note that proton-proton $g(r)$ changes with the
helium concentration but the correlation of helium nuclei does
not. With increasing presence of helium, a peak in the proton-proton
$g(r)$ appears at $r=1.4\,a_0$ which indicates the formation of H$_2$
molecules. Adding helium nuclei leads to the localization of a
fraction of the electrons. The available space in combination with the
reduced electronic exchange effects then leads to the formation of hydrogen
molecules which are not present in pure hydrogen at the same
temperature and density.

\section{CONCLUSIONS}

The discussion in this article centered on the phase diagram of hot
dense hydrogen. The first PIMC results for hydrogen-helium mixtures
were presented here, and the accuracy of existing helium EOS models as
well as the commonly used ideal mixing rule were analyzed. This
analysis revealed substantial inaccuracies in existing EOS
models. These uncertainties prevent us from making reliable models for
the interior of giant planets and from drawing conclusions about their
formation mechanism, either nucleated capture of nebular
hydrogen\cite{Li93} or gravitational instability driven formation in
protostellar disks.\cite{Boss97,Boss04}

An improved EOS will also yield a better characterization of the 120
extrasolar giant planets that have been discovered so far using radio
velocity measurements.\cite{Ma02} In particular the mass-radius
relationship obtained from transit extrasolar giant
planets\cite{brown01} will allow one to draw conclusions about the
heavy element abundance in those planets. The metallicity of a planet
relative to its central star constrains the role of accretion of
planetesimals in the planet's formation.

\section*{ACKNOWLEDGMENTS}

The author thanks W. Hubbard and D. Stevenson for useful discussions
and D. Saumon for providing us with the SCH EOS.


\end{document}